\documentclass[pmlr]{jmlr}

\usepackage{longtable}

\usepackage{booktabs}

\usepackage[load-configurations=version-1]{siunitx} 

\theorembodyfont{\upshape}
\theoremheaderfont{\scshape}
\theorempostheader{:}
\theoremsep{\newline}

\jmlrvolume{1}
\jmlryear{2024}
\jmlrworkshop{AI4Ed-AAAI}

\title[Improving Assessment of Tutoring Practices using Retrieval-Augmented Generation]{Improving Assessment of Tutoring Practices using Retrieval-Augmented Generation}

 \author{\Name{Zifei (FeiFei) Han} \Email{hanzifeifei@gmail.com}\\
\Name{Jionghao Lin} \Email{jionghao@cmu.edu}\\
\Name{Ashish Gurung} \Email{agurung@andrew.cmu.edu}\\
\Name{Danielle R. Thomas} \Email{drthomas@cmu.edu}\\
\Name{Eason Chen} \Email{easonc13@cmu.edu}\\
\Name{Conrad Borchers} \Email{cborcher@cs.cmu.edu}\\
 \Name{Shivang Gupta} \Email{Shivang@cmu.edu}\\
   \Name{Kenneth R. Koedinger} \Email{koedinger@cmu.edu}\\
   \addr Human-Computer Interaction Institute \\Carnegie Mellon University\\5000 Forbes Ave.\\Pittsburgh, PA 15213, USA}

\begin{document}

\maketitle

\begin{abstract}
One-on-one tutoring is an effective instructional method for enhancing learning, yet its efficacy hinges on tutor competencies. Novice math tutors often prioritize content-specific guidance, neglecting aspects such as social-emotional learning. Social-emotional learning promotes equity and inclusion and nurturing relationships with students, which is crucial for holistic student development. Assessing the competencies of tutors accurately and efficiently can drive the development of tailored tutor training programs. However, evaluating novice tutor ability during real-time tutoring remains challenging as it typically requires experts-in-the-loop. To address this challenge, this preliminary study aims to harness Generative Pre-trained Transformers (GPT), such as GPT-3.5 and GPT-4 models, to automatically assess tutors' ability of using social-emotional tutoring strategies. Moreover, this study also reports on the financial dimensions and considerations of employing these models in real-time and at scale for automated assessment. The current study examined four prompting strategies: two basic Zero-shot prompt strategies, Tree of Thought prompt, and Retrieval-Augmented Generator (RAG) based prompt. The results indicate that the RAG prompt demonstrated more accurate performance (assessed by the level of hallucination and correctness in the generated assessment texts) and lower financial costs than the other strategies evaluated. These findings inform the development of personalized tutor training interventions to enhance the the educational effectiveness of tutored learning.

\end{abstract}
\begin{keywords}
Large Language Model, Personalized Tutor Training, Automatic Assessment
\end{keywords}

\section{Introduction}
\label{sec:intro}

The efficacy of one-on-one tutoring for knowledge acquisition and retention continues to gather strong empirical evidence \citep{kraft2021blueprint, nickow2020impressive}. For tutoring to reach its optimal effectiveness, instructors need a multifaceted skill set, enabling them to provide both content-specific and social-emotional support to students \citep{thomas2023tutor, lin2023personalized, lin2022good, lin2023role, lin2022exploring}. However, it is often challenging for novice math tutors to blend both content-specific and social-emotional supports effectively into their teaching. They tend to focus primarily on content-specific instructional guidance, which can result in the inadvertent neglect of social-emotional learning components in their teaching methods \citep{inbook}. Social-emotional learning includes indispensable facets such as self-awareness, empathy, adeptness in building relationships, and critical decision-making abilities—crucial constituents for fostering the holistic development of students \citep{article}. Prior work shows that neglecting social-emotional learning during tutoring relates to significant loss in tutoring effectiveness \citep{marshall2021national}. Thus, assessing social-emotional learning support in tutors can guide personalized tutor training and subsequently improve tutoring effectiveness. However, this assessment has been hitherto expensive and infeasible for many educational applications, as it usually requires human experts, which are scarce \citep{kraft2021blueprint}.

In light of recent promising applications of large language models (LLMs) in education \citep{wang2023chatgpt, dai2023can, lin2023using}, this preliminary analysis aims to harness the potential of widely-used Generative Pre-trained Transformers (GPT), such as GPT-3.5 and GPT-4 to automatically assess social-emotional competencies in human tutors \citep{openai2023gpt4, lehman2022evolution}. We explore four types of prompting strategies: two basic Zero-shot prompts, the Tree of Thought prompt \citep{yao2023tree}, and a Retrieval-Augmented Generator (RAG)-based prompt \citep{lewis2020retrieval}. Additionally, given the significance of educational applications at scale, both in research and industry, it is crucial to understand the cost of using GPT models which is vital for assessing the economic feasibility of utilizing the model on our task. Consequently, this research investigates two \textbf{R}esearch \textbf{Q}uestions: \textbf{RQ1}: \textit{Can GPT models accurately assess the social-emotional learning competencies of human tutors?} \textbf{RQ2}: \textit{How does the performance and cost analysis of GPT-3.5 compare to that of GPT-4 in this context?}


\vspace{-3mm}

\section{Method}
\subsection{Data}


The dataset (tutoring dialogue transcripts) was collected from real-world middle-school math tutoring sessions in the United States. These sessions were conducted on Zoom, with novice human tutors to teach math ranging from Grade 6 to Grade 8. It should be noted that prior to the tutoring sessions, these human tutors completed some lessons involved social-emotional learning from a tutor training platform as described in \citet{lin2023personalized}. Each session, lasting approximately 30 minutes, involved dividing a group of students into multiple breakout rooms, assigning each student to an individual room. Tutors were responsible for managing three to five breakout rooms, conducting one-on-one tutoring as they circulated among them. When tutors noticed that certain students were capable of working independently with little assistance, they would move on to aid another student in a different room. The entire tutoring process was recorded via Zoom, and the recordings were transcribed using the speech-recognition tool Whisper \citep{radford2022robust}. In this work-in-progress study, we randomly selected five tutoring transcripts where a math tutor engages in tutoring sessions with five students, guiding them through the process of solving basic arithmetic problems.




\vspace{-3mm}


\subsection{Assessing Tutoring Practices}
\label{principles}

The tutoring dialogue transcripts were evaluated based on research-based principles in social-emotional learning and relationship building proposed by \citet{chhabra2022evaluation}. 
The details of these principles are presented in Appendix \ref{apd:third}; broadly we considered 5 categories: 
\textit{(1) Giving Effective Praise}, \textit{(2) Supporting a Growth Mindset}, \textit{(3) Reacting to Errors}, \textit{(4) Responding to Negative Self-Talk}, and \textit{(5) Using Motivational Strategies}. We used principles from these five most frequently used tutoring strategies to develop a rubric for assessing the tutoring practice.\footnote{The details of five lessons can be found via \url{https://www.tutors.plus/solution/training}} For brevity, we take the principle of \textit{Giving Effective Praise} as an example. 
This principle suggests that during tutoring, the instructor should focus on the praise on student learning effort instead of outcome. A desired praise is \textit{``You are almost there! I am proud of how you are persevering through and striving to solve the problem. Keep going!''} while an undesired praise is \textit{``You are so smart and almost got the problem correct.''} The principles and rubric from the five lessons further inform the design of prompt strategies to evaluate tutor's use of the practice. 

\vspace{-3mm}

\subsection{Large Language Model Generated Evaluation and Feedback}

To answer RQ1, we designed four types of prompt strategies. The rationale of prompt design follows two main elements where the prompt can foster GPT model to 1) generate score of the evaluation and 2) provide evidence from dialogue as the detailed interpretation of the score. By doing so, we designed four type of prompting strategies (see Appendix \ref{apd:first}) to analyze the the ability of GPT models on evaluating tutor performance, and these prompting strategies are detailed below:



\noindent\textbf{Basic Zero-shot Prompt Type I.} This zero-shot prompt was designed by using five effective tutoring principles of social-emotional learning and relationship building (detailed in the Appendix \ref{apd:first}). We prompt the GPT models to assess the whole tutoring transcript based on the tutoring principles in terms of scores and interpretation of score.


\noindent\textbf{Basic Zero-shot Prompt Type II.} This zero-shot prompt (detailed in the Appendix \ref{apd:first}) was used to firstly identify the incorrect use of tutoring practice from the transcript, and then assess the scores and provide interpretation of the scores.

\noindent\textbf{Tree of Thoughts (ToT) Prompt.}
As reported by \citet{yao2023tree}, the Tree of Thought prompt allows information to be organized in a structured way, like branches of a tree. This structured format helps the language model understand the information more comprehensively compared to simple and linear prompts. As a result, when using the Tree of Thought prompt, the language model might generate more precise and detailed responses. Our proposed ToT prompt (detailed in the Appendix \ref{apd:first}) is shown in \figureref{fig:imagetot}. The GPT models take the tutoring transcript as input. Then, the GPT models evaluate the tutor social-emotional learning based on the rubrics  from five principles (e.g., Using Motivation Strategies in \figureref{fig:imagetot}). As a result, the ToT-driven GPT models generate scores and interpretation of the scores on assessing the tutor social-emotional learning ability.

\begin{figure}[h]
\floatconts
  {fig:imagetot}
  {  \vspace{-5mm}\caption{Tree of Thought Evaluation Framework}}
  {\includegraphics[width=0.7\linewidth]{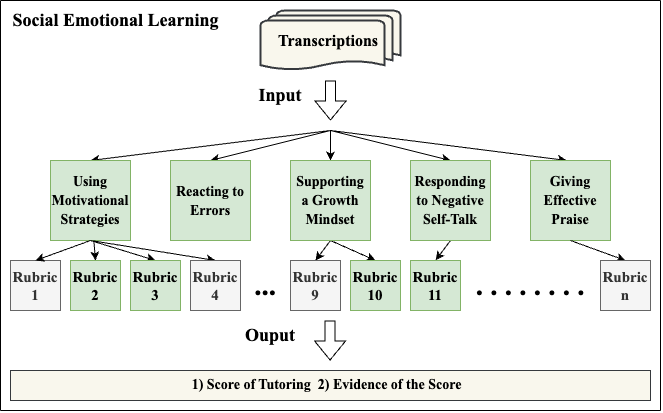}}
\end{figure}

\noindent\textbf{Retrieval-Augmented Generation (RAG).}
As reported by \citet{lewis2020retrieval}, RAG proves more effective than zero-shot prompts due to its integration of external knowledge sources (e.g., transcriptions and principles of tutoring), enriching the understanding of the GPT model and enabling it to generate more contextually relevant and accurate responses. Our study developed an RAG-based prompt (detailed in the Appendix \ref{apd:first}) as depicted in \figureref{fig:imageragengine}. Within the information database component (see \figureref{fig:imageragengine}), tutoring transcriptions and principles on social-emotional learning are initially converted into word embeddings (i.e., words represented as vectors for semantic relationships, as described in \citet{kusner2015word}) stored within the database. These embeddings, stored within information database, form the basis for our RAG-based prompt, allowing the GPT model to access a broader and more relevant set of information. In the actual evaluation process, the RAG model's evaluation engine (see \figureref{fig:imageragengine}) can selectively retrieve and incorporate information from our prepared word embeddings, guided by the principles of social-emotional learning. This selective retrieval ensures that the GPT model focuses on the most relevant aspects of the tutoring transcripts, ultimately enabling the RAG-based prompt to generate scores and provide evidence for these scores, illustrating the potential of RAG in enhancing GPT models.

\begin{figure}[htbp!]
\floatconts
  {fig:imageragengine}
  {\caption{The structure of Retrieval-Augmented Generation (RAG) based prompt}}
  {\includegraphics[width=0.75\linewidth]{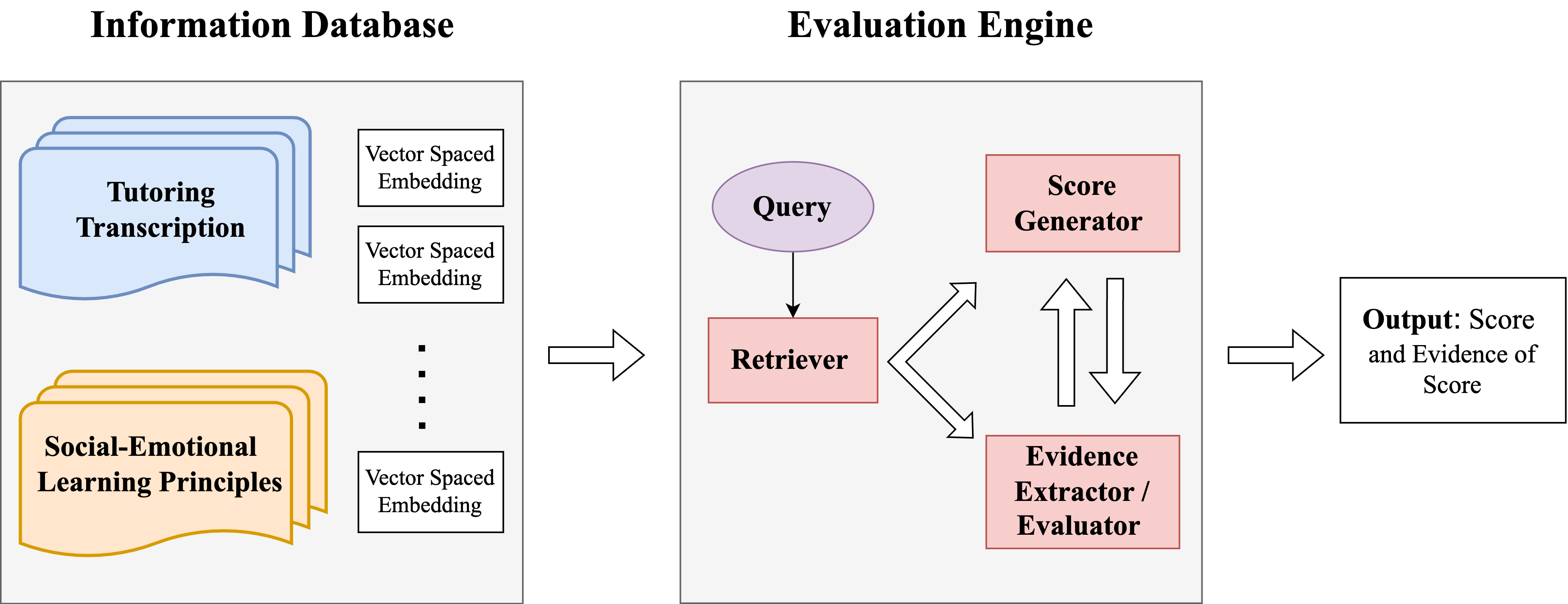}}
\end{figure}


\subsection{Evaluation Metrics}
\label{metrics}


In this work-in-progress paper, annotation was conducted by a single human coder since we aim to establish a preliminary understanding of evaluating the accuracy of GPT models' output, which can be later expanded and refined in subsequent phases of research. While the use of one coder suits our current exploratory needs, we recognize its limitation in inter-rater reliability. To address this, we plan to involve multiple coders in future stages of the research, which will enable us to conduct a thorough inter-rater reliability analysis and ensure the robustness of our findings. The human coder annotated the GPT generated output based on two metrics: 1) \textbf{Correctness} and 2) \textbf{Hallucination}, which are detailed below: 
\begin{itemize}

    \item  \textbf{Correctness} is the metric to evaluate the capability of GPT models in accurately assessing the tutor's use of social-emotional learning principles within the tutoring transcript. Evaluating correctness is essential for verifying the model's understanding and interpretation of the given information. Human coders review the GPT-generated assessment results, which include feedback and scores, against the actual tutoring transcript. The scoring categories for this metric include \textit{``-1''} indicating that no information was generated for a specific social-emotional learning principle, \textit{``0''} for GPT model incorrect assessment, and \textit{``1''} for GPT model correct assessment. The correctness is guided by a rubric based on the five principles described in Section. \ref{principles}.

    \item  \textbf{Hallucination} denotes the phenomenon where a generative model (e.g., GPT models) produces content that deviates from factual accuracy or logical coherence with respect to the input prompt or source content \cite{ji2023survey}. Hallucinations may manifest as fabricated facts, illogical statements, or irrelevant responses that do not align with the established context or contradict the known data \cite{ji2023survey}.  In our study, we noted instances where the GPT model generated feedback that was unrelated to events in the tutoring transcript, exemplifying the issue of hallucination. This issue poses significant challenges for applications that rely on the veracity of generated text, necessitating rigorous validation mechanisms to ensure the reliability of model outputs in critical domains. To measure the hallucination of generated text, the human coder annotated the output into \textit{``-1''}, \textit{``0''}, \textit{``0.5''}, and \textit{``1''} where \textit{``-1''} indicates that no information was generated for a specific social-emotional learning principle; \textit{``0''} indicates no hallucination; \textit{``0.5''} signifies partial hallucination (both hallucinated and non-hallucinated information coexist in the text), and \textit{``1''} represents a completely hallucinated response.

\end{itemize}


\noindent\textbf{Evaluation on Financial Cost.} To answer RQ2, we recorded the cost for using different GPT models with different prompts. We counted the input token from prompt and transcript whereas the output tokens from GPT generated text. We called the API of GPT-3.5 Turbo and GPT-4 Turbo. By referring the GPT API price,\footnote{\url{https://openai.com/pricing}} we calculated the cost associated with each generated text from GPT models.

\section{Results}

\begin{figure}[htbp!]
\floatconts
  {fig:image}
  {\caption{Analysis of GPT models' accuracy across various prompt strategies. Zero-shot P1 and P2 denote the basic Zero-shot prompt type I and type II, respectively}}
  {\includegraphics[width=0.85\linewidth]{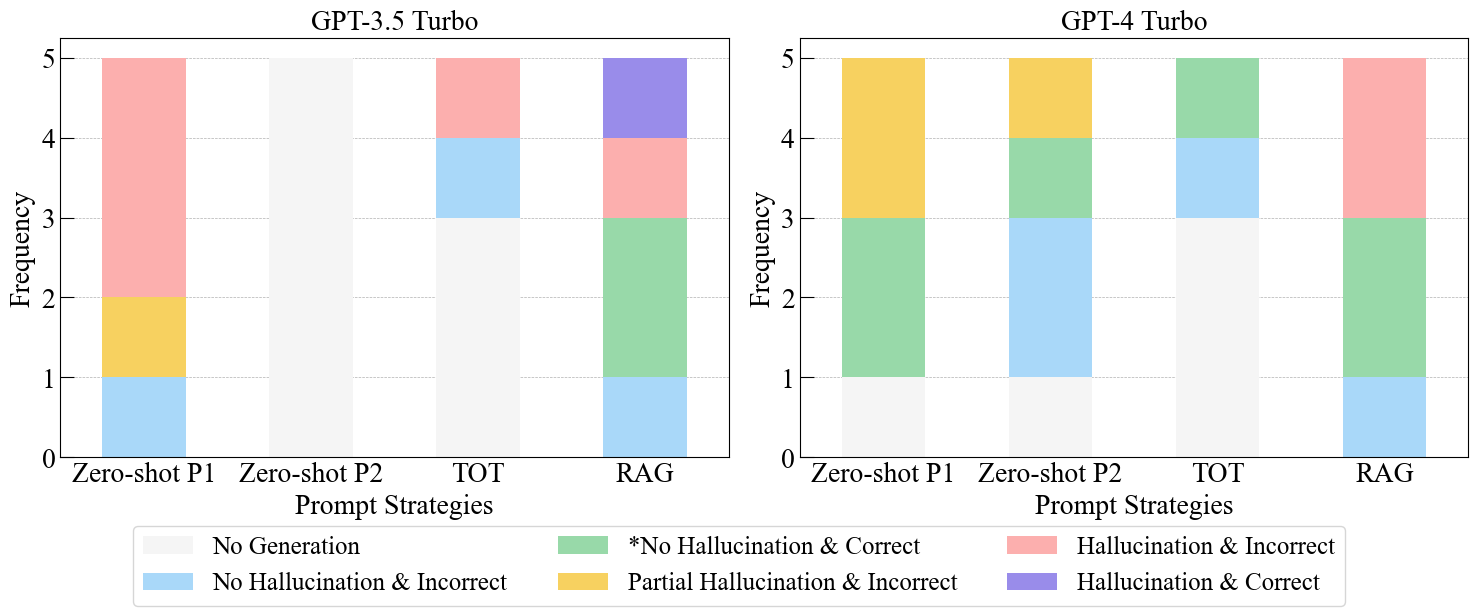}}
  \label{result:rq1}
\end{figure}



\noindent\textbf{Results on RQ1.} We present the results in Figure \ref{result:rq1}, which illustrate the accuracy of GPT models on evaluating the tutoring transcript across various prompt strategies. The accuracy of GPT models was measured by hallucination and correctness metrics (detailed in Method \ref{metrics}), which aimed to assess GPT models' ability to evaluate a tutor's competency in social-emotional learning. Notably,  the green area in Figure \ref{result:rq1} represents instances of no hallucination and correct evaluation, which is the desired evaluation on the tutoring transcript. The left side of Figure \ref{result:rq1} showed the accuracy of GPT-3.5 model. This demonstrates that the RAG strategy has the potential in providing no hallucination and correct evaluation of the transcripts. In comparison, other prompting strategies from GPT-3.5 failed to provide desired evaluation on the tutoring transcript. It should be noted that the results from GPT-3.5 include an evaluation stating, \textit{``The tutor did not respond to negative self-talk by validating the student's feelings or building their self-efficacy.''} On closer examination of the tutoring transcript, we observed that the student did not engage in negative self-talk, hence the tutor's lack of response. However, the GPT-3.5 evaluation implied that the tutor intentionally did not respond to negative self-talk. Thus, our human annotator identified as a hallucination. In contrast, the GPT-4 model results (on the right in Figure \ref{result:rq1}) show a generally more accurate evaluation (evidenced by more frequent green area) compared to the GPT-3.5 model. This suggests the advanced capability of GPT-4, potentially outperforming the GPT-3.5 model. It is important to note that both the RAG and Zero-shot (P1) prompts in GPT-4 yielded more accurate evaluations than other prompting strategies. The comparative analysis of GPT-3.5 and GPT-4 underscores that the RAG-based prompting strategy consistently produces the desired output, highlighting its effectiveness across different model versions.

\noindent\textbf{Results on RQ2.} In our subsequent analysis, we evaluated the financial cost per lesson evaluation for both GPT-3.5 Turbo and GPT-4 Turbo using various prompting strategies, as outlined in Table \ref{result:rq2}. This analysis indicated that the financial cost associated with GPT-4 is approximately 10 times greater than that of GPT-3.5. The Retrieval-Augmented Generation (RAG)-based prompting strategy was identified as the most cost-effective for both the GPT-3.5 and GPT-4 models. As discussed earlier in the Results section on Research Question 1 (RQ1), the RAG-based approach has demonstrated its capability to provide accurate evaluations without hallucinations in tutoring transcripts for both models.  Consequently, we propose the RAG-based prompt as the most cost-efficient strategy. These findings provide insights into the selection of different prompting strategies and GPT models, particularly in terms of balancing effectiveness with financial constraints.

\begin{table}[hbt!]
\footnotesize
\caption{Comparison of costs across various prompting strategies in GPT-3.5 and GPT-4}
\label{result:rq2}
\renewcommand{\arraystretch}{1.3}
\begin{tabular}{@{}p{8cm}m{3cm}m{3cm}@{}}
\hline
\textbf{Prompt}                                             & \textbf{GPT-3.5 Turbo} & \textbf{GPT-4 Turbo}   \\ \hline 
                                                                          
\textbf{\begin{tabular}[c]{@{}l@{}}Zero-shot Prompt Type I\end{tabular}}          & \$0.100                & \$1.035       \\ \hline
\textbf{\begin{tabular}[c]{@{}l@{}} Zero-shot Prompt Type II \end{tabular}} & \$0.014                & \$0.188       \\ \hline
\textbf{\begin{tabular}[c]{@{}l@{}}Tree of Thoughts (ToT)\end{tabular}} & \$0.013                & \$0.137       \\ \hline
\textbf{\begin{tabular}[c]{@{}l@{}}Retrieval-Augmented Generation (RAG)\end{tabular}} & \$0.008 & \$0.137 \\ \hline
\end{tabular}
\end{table}

\section{Conclusion}
This preliminary study highlights the potential of Retrieval-Augmented Generation (RAG) prompting in evaluating the quality of tutoring based on social-emotional learning competencies. By integrating tutoring transcripts and principles into the GPT model via word embeddings, as elaborated in Table \ref{tab:rag} in our appendix, RAG enables more contextually relevant and precise evaluations. The RAG-based prompt not only showcased more accurate performance in evaluating tutoring practices and lower financial costs compared to other prompts but also laid the groundwork for broader applications in tutor skill assessment. Moving forward, we aim to further explore the RAG prompt's effectiveness in assessing additional tutoring competencies, such as building content skills and promoting inclusion. Additionally, we plan to employ the RAG prompt to evaluate a broader range of real-world tutoring transcripts, assessing its efficacy across diverse tutoring interactions. Our goal is to identify areas where tutors may lack skills and provide corresponding training lessons to help them improve. As an extension of this study, we intend to design a training lesson recommender system that can process the assessment from the GPT model and offer lesson recommendations to assist tutors in enhancing their tutoring skills. A prototype developed for the demonstration of the lesson recommendation system is accessible at \url{https://tutorevaluation.vercel.app}.

\bibliography{pmlr-sample}

\newpage 
\appendix

\section{Prompting Strategies}\label{apd:first}

\begin{table}[hbt!]
\footnotesize
\caption{Basic Zero-shot Prompt Type I}
\label{tab:standard}
\begin{tabular}{@{}p{1.5cm}m{13cm}@{}}
\hline
& \textbf{Prompt}  \\ \hline
\textbf{Scoring}   & \textit{Given a dialogue of a tutoring session, please evaluate the Tutor based on specific best teaching practice within \texttt{\{Principle\_Name\}} Please return 1 if the tutor correctly used the tutoring practice. 
Return 0 if the tutor incorrectly used the tutoring practice. 
Please only return 0 or 1} \\ \hline
\textbf{Generator} & \textit{Please briefly explain why you give the score?}   \\ \hline
\end{tabular}
\end{table}

\begin{table}[ht!]
\footnotesize
\caption{Basic Zero-shot Prompt Type II}
\label{tab:IDprompting}
\begin{tabular}{@{}p{1.5cm}m{13cm}@{}}
\hline
                         & Prompt                                                                                                                                                                                                                                                                                                                                                                                                                                                                                                                     \\ \hline
\textbf{Incorrect Identification} & \textit{Given the following evaluation criteria \texttt{\{Principle\_Criteria\}} Please identify if there is any tutor's incorrect use of the tutoring strategy \texttt{\{Principle\_Name\}}based on the criteria above.If there is incorrect response, return the incorrect response by tutor as evidence in the dialogue and list the criteria not met. If the tutor used teaching strategy correctly, please return based on the criteria above, which ones are correct and also return evidence from the dialogue.}  \\ \hline
\textbf{Score Generation}   & \textit{Return the score of \texttt{\{Principle\_Name\}} from 0 to 5 based on the evaluation. Give one point to each criteria met.}                                                                                                                                                                                                                                                                                                                                                                                                                  \\ \hline
\end{tabular}
\end{table}

\begin{table}[hbt!]
\footnotesize
\caption{Tree of Thought Prompt}
\label{tab:tot}
\begin{tabular}{@{}p{1.5cm}m{13cm}@{}}
\hline
                      & \textbf{Prompt}        \\ \hline
\textbf{Layer\_1} 
& \textit{\texttt{\{Social\_Emotional\_Learning\_Principles\}} For the following transcript between a tutor and a middle school student, score how well the tutor performed in the competency area above. Give one point for each of the following criteria or skills being met by the tutor. For example, if a tutor did not demonstrate any evidence of a given skill or criteria give a score of 0.  If a tutor met all the given criteria, give a score of 5. Please only return the evaluated score from 0 to 5.} \\ \hline
\textbf{Layer\_2} 
& \textit{For each criteria listed, please indicate which from the current \texttt{\{Social\_Emotional\_Learning\_Principles\}} is not met, and which criteria are met.}                                                                                                                                                                                                                                                                                                                                                                                                      \\ \hline
\textbf{Layer\_3}
& \textit{Given a dialogue of a tutoring session between a tutor and a middle school student, please evaluate the Tutor based on specific given criteria : \texttt{\{rubric\}}, Please return 1 if the tutor correctly used the tutoring practice.
Return 0 if the tutor incorrectly used the tutoring practice. Provide your evaluation in the form of a number. Please also list evidence why you provide the evaluation.}                                                                                                                                                                  \\ \hline
\end{tabular}
\end{table}

\begin{table}[hbt!]
\footnotesize
\caption{Retrieval-Augmented Generation Based Prompt}
\label{tab:rag}
\begin{tabular}{@{}p{1.5cm}m{13cm}@{}}
\hline
          & \textbf{Prompt}                                                                                                                                                                                                                                                                                                                                                    \\ \hline
\textbf{Retriever} & \textit{For each criteria and rubric above, please identify all of tutor's correct and incorrect use of the practice above.  Return the dialogues of the tutor as evidence in the format: 1. Competency 2. Each Criteria of the competency  3. Sentences that tutor said within the dialogue serves as evidence.}
\\ \hline
\textbf{Generator} & \textit{Return the score of \texttt{\{Principle\_Name\}} from 0 to 5 based on the evaluation. Give one point to each criteria met. Please only return the evaluated score from 0 to 5.}                          \\ \hline
\end{tabular}
\end{table}

\clearpage 

 \section{Social-Emotional Learning Principles}\label{apd:third}

\begin{table}[hbt!]
\footnotesize
\caption{Social Emotional Learning Principles}
\label{tab:lessons}
\begin{tabular}{@{}p{3cm}m{11cm}@{}}
\hline
\textbf{Principles}                           & \textbf{Description}                                                                                                                                                                                                            \\ \hline
\textbf{Giving Effective Praise}          & Praising students for putting forth effort by giving process-focused praise instead of praising students for getting an answer correct or getting a good grade                                                                  \\ \hline
\textbf{Supporting a Growth Mindset}      & Supporting a growth mindset instead of a fixed mindset by encouraging students on the learning process and not necessarily just getting the answer                                                                              \\ \hline
\textbf{Reacting to Errors}               & Responding to students when students make errors or mistakes, by not directly calling attention to the error but guiding students to realize and correct the error themselves.                                                  \\ \hline
\textbf{Responding to Negative Self-Talk} & Responding to students positively when students engage in negative self-talk, such as saying ``I can't do this'' or ``this is too hard for me'' by validating a student’s feelings but encouraging and building their self-efficacy \\ \hline
\textbf{Using Motivational Strategies}    & Rewarding students by using intrinsic and extrinsic motivation strategies, such as rewarding students for working hard by giving them time at the end of a session to discuss their interests                                   \\ \hline
\end{tabular}
\end{table}

\end{document}